\documentclass[aps,preprint,floatfix,10point]{revtex4}

\usepackage{graphicx,color}
\usepackage{psfrag}
\usepackage{amssymb}
\usepackage{eso-pic}
\usepackage[dvips,letterpaper]{geometry} 
\usepackage{amsmath}

\begin{document}

%
%
\def\oti{{\otimes}}
\def\lb{ \left[ }
\def\rb{ \right]  }
\def\tilde{\widetilde}
\def\bar{\overline}
\def\hat{\widehat}
\def\*{\star}
\def\[{\left[}
\def\]{\right]}
\def\({\left(}		\def\BL{\Bigr(}
\def\){\right)}		\def\BR{\Bigr)}
	\def\BBL{\lb}
	\def\BBR{\rb}
%
%
\def\zb{{\bar{z} }}
\def\zbar{{\bar{z} }}
\def\frac#1#2{{#1 \over #2}}
\def\inv#1{{1 \over #1}}
\def\half{{1 \over 2}}
\def\d{\partial}
\def\der#1{{\partial \over \partial #1}}
\def\dd#1#2{{\partial #1 \over \partial #2}}
\def\vev#1{\langle #1 \rangle}
\def\ket#1{ | #1 \rangle}
\def\rvac{\hbox{$\vert 0\rangle$}}
\def\lvac{\hbox{$\langle 0 \vert $}}
\def\2pi{\hbox{$2\pi i$}}
\def\e#1{{\rm e}^{^{\textstyle #1}}}
\def\grad#1{\,\nabla\!_{{#1}}\,}
\def\dsl{\raise.15ex\hbox{/}\kern-.57em\partial}
\def\Dsl{\,\raise.15ex\hbox{/}\mkern-.13.5mu D}
%
%
\def\ga{\gamma}		\def\Ga{\Gamma}
\def\be{\beta}
\def\al{\alpha}
\def\ep{\epsilon}
\def\vep{\varepsilon}
\def\la{\lambda}	\def\La{\Lambda}
\def\de{\delta}		\def\De{\Delta}
\def\om{\omega}		\def\Om{\Omega}
\def\sig{\sigma}	\def\Sig{\Sigma}
\def\vphi{\varphi}
%
%
\def\CA{{\cal A}}	\def\CB{{\cal B}}	\def\CC{{\cal C}}
\def\CD{{\cal D}}	\def\CE{{\cal E}}	\def\CF{{\cal F}}
\def\CG{{\cal G}}	\def\CH{{\cal H}}	\def\CI{{\cal J}}
\def\CJ{{\cal J}}	\def\CK{{\cal K}}	\def\CL{{\cal L}}
\def\CM{{\cal M}}	\def\CN{{\cal N}}	\def\CO{{\cal O}}
\def\CP{{\cal P}}	\def\CQ{{\cal Q}}	\def\CR{{\cal R}}
\def\CS{{\cal S}}	\def\CT{{\cal T}}	\def\CU{{\cal U}}
\def\CV{{\cal V}}	\def\CW{{\cal W}}	\def\CX{{\cal X}}
\def\CY{{\cal Y}}	\def\CZ{{\cal Z}}
\def\rvac{\hbox{$\vert 0\rangle$}}
\def\lvac{\hbox{$\langle 0 \vert $}}
\def\comm#1#2{ \BBL\ #1\ ,\ #2 \BBR }
\def\2pi{\hbox{$2\pi i$}}
\def\e#1{{\rm e}^{^{\textstyle #1}}}
\def\grad#1{\,\nabla\!_{{#1}}\,}
\def\dsl{\raise.15ex\hbox{/}\kern-.57em\partial}
\def\Dsl{\,\raise.15ex\hbox{/}\mkern-.13.5mu D}
%
%
\font\numbers=cmss12
\font\upright=cmu10 scaled\magstep1
\def\stroke{\vrule height8pt width0.4pt depth-0.1pt}
\def\topfleck{\vrule height8pt width0.5pt depth-5.9pt}
\def\botfleck{\vrule height2pt width0.5pt depth0.1pt}
\def\Zmath{\vcenter{\hbox{\numbers\rlap{\rlap{Z}\kern
0.8pt\topfleck}\kern 2.2pt
                   \rlap Z\kern 6pt\botfleck\kern 1pt}}}
\def\Qmath{\vcenter{\hbox{\upright\rlap{\rlap{Q}\kern
                   3.8pt\stroke}\phantom{Q}}}}
\def\Nmath{\vcenter{\hbox{\upright\rlap{I}\kern 1.7pt N}}}
\def\Cmath{\vcenter{\hbox{\upright\rlap{\rlap{C}\kern
                   3.8pt\stroke}\phantom{C}}}}
\def\Rmath{\vcenter{\hbox{\upright\rlap{I}\kern 1.7pt R}}}
\def\Z{\ifmmode\Zmath\else$\Zmath$\fi}
\def\Q{\ifmmode\Qmath\else$\Qmath$\fi}
\def\N{\ifmmode\Nmath\else$\Nmath$\fi}
\def\C{\ifmmode\Cmath\else$\Cmath$\fi}
\def\R{\ifmmode\Rmath\else$\Rmath$\fi}

\def\barray{\begin{eqnarray}}
\def\earray{\end{eqnarray}}
\def\beq{\begin{equation}}
\def\eeq{\end{equation}}

\def\n{\noindent}

\def\Tr{\rm Tr} 
\def\xvec{{\bf x}}
\def\kvec{{\bf k}}
\def\kvecp{{\bf k'}}
\def\omk{\om{\kvec}} 
\def\dk#1{\frac{d\kvec_{#1}}{(2\pi)^d}}
\def\2pid{(2\pi)^d}
\def\ket#1{|#1 \rangle}
\def\bra#1{\langle #1 |}
\def\vol{V}
\def\adag{a^\dagger}
\def\rme{{\rm e}}
\def\Im{{\rm Im}}
\def\pvec{{\bf p}}
\def\fermiS{\CS_F}
\def\cdag{c^\dagger}
\def\adag{a^\dagger}
\def\bdag{b^\dagger}
\def\vvec{{\bf v}}
\def\muhat{{\hat{\mu}}}
\def\vac{|0\rangle}
\def\pcut{{\Lambda_c}}
\def\chidot{\dot{\chi}}
\def\gradvec{\vec{\nabla}}
\def\psitilde{\tilde{\Psi}}
\def\psibar{\bar{\psi}}
\def\psidag{\psi^\dagger} 
\def\m{m_*}
\def\up{\uparrow}
\def\down{\downarrow}
\def\Qo{Q^{0}}
\def\vbar{\bar{v}}
\def\ubar{\bar{u}}
\def\smallhalf{{\textstyle \inv{2}}}
\def\smallsqrt{{\textstyle \inv{\sqrt{2}}}}
\def\rvec{{\bf r}}
\def\avec{{\bf a}}
\def\pivec{{\vec{\pi}}}
\def\svec{\vec{s}} 
\def\phivec{\vec{\phi}}
\def\daggerc{{\dagger_c}}
\def\Gfour{G^{(4)}}
\def\dim#1{\lbrack\!\lbrack #1 \rbrack\! \rbrack }
\def\qhat{{\hat{q}}}
\def\ghat{{\hat{g}}}
\def\nvec{{\vec{n}}}
\def\bull{$\bullet$}
\def\ghato{{\hat{g}_0}}
\def\r{r}
\def\deltaq{\delta_q}
\def\gcharge{g_q}
\def\gspin{g_s}
\def\deltas{\delta_s}
\def\gQC{g_{AF}} 
\def\ghatqc{\ghat_{AF}}
\def\xqc{x_{AF}}
\def\mhat{\hat{m}}
\def\xup{x_2}
\def\xdown{x_1}
\def\sigmavec{\vec{\sigma}}
\def\xopt{x_{\rm opt}}
\def\Lambdac{{\Lambda_c}}
\def\angstrom{{{\scriptstyle \circ} \atop A}     }
\def\AA{\leavevmode\setbox0=\hbox{h}\dimen0=\ht0 \advance\dimen0 by-1ex\rlap{
\raise.67\dimen0\hbox{\char'27}}A}
\def\ratio{\gamma}
\def\Phivec{{\vec{\Phi}}}
\def\singlet{\chi^- \chi^+} 
\def\mhat{{\hat{m}}}

\def\Im{{\rm Im}}
\def\Re{{\rm Re}}

\def\xstar{x_*}
\def\sech{{\rm sech}}
\def\Li{{\rm Li}}
\def\dim#1{{\rm dim}[#1]}
\def\ep{\epsilon}
\def\free{\CF}
\def\Fhat{\digamma}
\def\ftilde{\tilde{f}}
\def\muphys{\mu_{\rm phys}}
\def\xitilde{\tilde{\xi}}
\def\CI{\mathcal{I}}
\def\nhat{\hat{n}}
\def\ef{\epsilon_F}
\def\as{a_s}
\def\diffk{|\kvec - \kvec' |}
\def\bfT{{\bf T}}

\def\bfC{{\bf C}}
\def\bfP{{\bf P}}

\def\Ct{ \tilde{C} }
\def\Tt{ \tilde{T} }
\def\Pt{ \tilde{P} }
\def\etatilde{\tilde{\eta}}
\def\pt{\tilde{p}}
\def\Ttilde{\Tt}
\def\Ctilde{\Ct}
\def\Ptilde{\Pt}

\def\zbar{{\overline{z}}}
\def\dbar{{\overline{d}}}
\def\pbar{{\overline{\d }}}
\def\Abar{{\overline{A}}}

\def\To{T_o}
\def\Ttildeo{\Ttilde_o}
\def\Te{T_e}
\def\Ttildee{\Ttilde_e}

\def\Co{C_o} 
\def\Ctildeo{\Ctilde_o}
\def\Ce{C_e} 
\def\Ctildee{\Ctilde_e}

\def\Z{\mathbb{Z}}

\def\none{$\emptyset$}

\def\Ctilde{\tilde{C}}
\def\one{{\bf 1}}

\def\Gtilde{\tilde{G}}


\title{Holographic classification of Topological Insulators and  its 8-fold periodicity}
\author{ Andr\'e  LeClair$^{\clubsuit}$  and Denis Bernard$^\spadesuit$}
\affiliation{ $^\clubsuit$Cornell University, Ithaca, NY\\
\centerline{$^\spadesuit$Laboratoire de physique th\'eorique, Ecole Normale Sup\'erieure,  CNRS, Paris, France}\\ 
  }

\vskip 2.5 truecm

\begin{abstract}

Using generic properties of Clifford algebras in any spatial dimension,   we  explicitly  classify Dirac 
hamiltonians with zero modes protected by the discrete symmetries of time-reversal,  particle-hole
symmetry, and chirality.    Assuming the boundary states of topological insulators are Dirac fermions, 
  we  thereby  holographically  reproduce the Periodic Table of topological insulators 
found by Kitaev\cite{Kitaev} and Ryu. et. al.\cite{Ryu},    without using topological invariants nor
K-theory.  In addition we find candidate  $\Z_2$  topological insulators in classes  AI, AII in dimensions 0,4  mod 8  and in 
classes C, D in dimensions 2,6  mod 8.

\end{abstract}

\maketitle

\section{Introduction}

Topological insulators are characterized by bulk band structures with special topological properties\cite{Kane,Zhang,Fu,Fu2,Moore,Konig,Hsieh}.
Namely,  from the bulk wave-functions in momentum space,  one can construct a gauge field and the topological invariant is 
essentially a Chern number.   These physical systems possess a 
bulk/boundary correspondence,  in that they necessarily have protected gapless excitations
on the $\dbar = d-1$ dimensional surface.    These surface modes are typically described by
Dirac hamiltonians.   For example in the integer quantum Hall effect (QHE)  in $d=2$,  the Chern number  is
the same integer as  in the quantized Hall conductivity, and the edge states are chiral Dirac fermions\cite{Thouless,Stone}.  

 Kitaev\cite{Kitaev}   and Ryu.  et. al\cite{Ryu}   classified topological insulators (TI)  in any spatial  dimension 
according to the discrete symmetries  of time reversal $\bfT$,  particle-hole symmetry $\bfC$ and 
chirality $\bfP$ and  found 5 classes of topological insulators in any dimension.    These classifications 
were based on the existence of topological invariants\cite{Ryu}  or   K-theory\cite{Kitaev}.   
The bulk/boundary correspondence was pointed out in \cite{Schnyder}  for $d=3$ spatial dimensions:   using the classification 
of  $\dbar =d-1 =2$ dimensional Dirac hamiltonians in \cite{BL},  it was found that precisely 5 of the 13  Dirac classes
had protected zero modes  with the predicted  discrete symmetries.    In that analysis,  it was crucial that
the classification in \cite{BL} contained  3 additional classes beyond the 10 Altland-Zirnbauer (AZ) classes,
since it was precisely these additional classes that corresponded to some of the topological insulators.   

This ``holographic''   classification of topological insulators,  i.e. based on the existence of symmetry-protected 
zero modes on the boundary,   is not necessarily equivalent to a classification based on topology or K-theory.    Indeed,  
this issue was studied in \cite{EunAh} for $d=2$,  and 6 additional possible classes of 
TI's were found in addition to the predicted 5.         This motivated the present work,  which presents  a holographic classification of TI's in
any dimension.  We should emphasize from the beginning that we do not perform a complete classification of the most general 
Dirac hamiltonian in any dimension,  as was done in \cite{BL,EunAh}  for $\dbar =2,1$,  since this would hardly be useful in higher dimensions. 
Rather,   the  goal is to begin with a minimal form of Dirac hamiltonian,  and then classify those which have protected zero modes.   
If one can thereby  reproduce the Periodic Table of TI's as given in \cite{Ryu,Kitaev},   then this supports the validity of  this holographic classification
and further supports the $d=2$ results in \cite{EunAh}.     Our approach is closest to Kitaev's,  since here the 8-fold periodicity arises from
the well known 8-fold periodicity of  the reality properties of spinor representations of the orthogonal groups,  which is a mild form of
Bott periodicity in K-theory;   however no K-theory arguments are invoked in the present  work.  
We also wish to emphasize that in the present work,  our construction is based only on  the {\it generic}  properties of Dirac operators in
any dimension $d$,  and thus,  as expected,   we do not find all of the exceptional cases in $d=2$ found in\cite{EunAh}.

The remainder of the paper is organized as follows.   In the next section we review the definitions of the 10 Altland-Zirnbauer (AZ)   classes.    
In section III we formulate the classification problem in terms of Clifford algebras.      The specific representation of  the Clifford algebras we will
use are presented in section IV.    In section V  we describe  how to  realize  all 10  AZ  classes in any dimension.    Section VI  contains our classification
of protected zero modes;   we reproduce the Periodic Table,  and find one additional candidate for a TI in every even dimension.   
In section VII we briefly describe why $d=2$ is exceptional.

\section{Discrete symmetries}

The  10 Altland-Zirnbauer  (AZ) classes of random hamiltonians   arise when one considers 
time reversal symmetry $(\bfT)$,  particle-hole symmetry $(\bfC)$, and parity or chirality $(\bfP)$\cite{AZ}.   
These discrete symmetries are defined  to act as follows on a first-quantized hamiltonian $\CH$:
\barray
\nonumber 
\bfT:  ~~~~~~~~~~T \CH^* T^\dagger &=& \CH
\\ 
\bfC: ~~~~~~~~~C \CH^T C^\dagger &=& - \CH 
\label{syms} 
\\ 
\bfP: ~~~~~~~ ~~ ~~P \CH P^\dagger &=& - \CH \nonumber 
\earray
with $T T^\dagger = C C^\dagger = P P^\dagger ={\bf 1} $,  and $\CH^T$ denotes the transpose of $\CH$.    
In our classification,  two hamiltonians $\CH, \CH'$ related by a unitary transformation $\CH' = U \CH U^\dagger$ are in the same class,  since they have the same eigenvalues.   For $C$ and $T$,  this translates to $C \to C' = UCU^T$  and $T \to T' = UTU^T$.    For $P$,  the unitary   transformation is $P \to P'=U P U^\dagger$. 
In the sequel,  we will refer to these unitary transformations as gauge transformations.  

For hermitian hamiltonians, $\CH^T = \CH^*$,  thus, up to a sign, $\bfC$ and $\bfT$ symmetries are the same. We focus then on these symmetries
involving the transpose:   $T \CH^T  T^\dagger =  \CH$ and $C \CH^T C^\dagger = - \CH$.   Taking the transpose of this relation, one finds there are two consistent possibilities: $T^T = \ep_t  T$ and $ C^T = \ep_c  C$, where $\ep_{t,c} = \pm 1$,   which are gauge-invariant relations.
   The various classes are  thus distinguished by $\ep_t  = \pm 1, \emptyset$ and $\ep_c = \pm 1, \emptyset$,    where $\emptyset$ indicates that the hamiltonian does not have the symmetry.
   (In some literature,  $T,C$ are chosen to be real,  unitarity implies $T^2 = \ep_t$, $C^2 = \ep_c$, and this sign of the square 
   characterizes the classes;   however this is not a gauge-invariant statement.)
 One obtains $9 = 3 \times 3$ classes just by considering the 3 cases for $\bfT$ and $\bfC$.    If the hamiltonian has both $\bfT$ and $\bfC$ symmetry, then it automatically has a $\bfP$ symmetry, with $P=TC^\dagger$ up to a phase.  If there is neither $\bfT$ nor
  $\bfC$  symmetry,  then there are two choices $P=\emptyset,1$,  and this gives the  additional class AIII,  leading to a total of 10.   Their properties are shown in 
  Table \ref{Table1}. We also mention that one normally requires $P^2 =1$.  Below, we will require $\bfT$ and $\bfC$ to commute,  thus $P^2 = T^2 { C^\dagger}^2 = \pm 1$.  However one has the freedom $P \to i P$ to restore $P^2 =1$.    In the sequel,  in the cases with both $\bfT, \bfC $ symmetry,  we simply define $P = TC^\dagger$, up to a phase.  

\begin{table}
\begin{center}
\begin{tabular}{|c|c|c|c|}
\hline\hline
AZ-classes &  ~ ~~$T$~~ ~ &~~~ $C$~ ~~&~~~ $P$ ~ ~~ \\
\hline\hline 
A  & \none  & \none & \none \\
AIII   & \none & \none & 1  \\
AII  &$ -1 $& \none &\none \\
AI & +1 &\none&\none \\
C & \none & $-1$&\none \\
D  & \none & +1 & \none  \\
BDI   &+1 & +1 & 1  \\
DIII   & $-1$ & +1 & 1  \\
CII &$ -1$ &$ -1$ & 1 \\
CI & +1 &$ -1$ & 1 \\
\hline\hline 
\end{tabular}
\end{center}
\caption{\emph{The 10 Altland-Zirnbauer (AZ) hamiltonian classes.  The $\pm$ signs refer to
$T^T = \pm T$ and $C^T= \pm C$,  whereas \none ~  denotes non-existence of the symmetry. }}
\label{Table1}
\end{table}

\def\utilde{\tilde{u}}

\section{Formulation in terms of Clifford algebras}

Let $d$ denote the spatial dimension and $\dbar = d-1$ the dimension of the boundary.    On the boundary,
we assume a  first quantized Dirac hamiltonian of the form:
\beq
\label{ham}
\CH  =  - i \sum_{a=1}^\dbar  \gamma_a  \frac{ \d}{\d{x_a}}  +  M 
\eeq
where $x_a$ are coordinates on the boundary and $\gamma_a ,  M$ are matrices.   
In momentum space $\kvec$,  in order for the hamiltonian to satisfy $\CH^2 =  \kvec^2 + M^2$,  
and have a single particle energy spectrum $E= \pm  \sqrt{\kvec^2 + M^2}$, 
the $\gamma_a$,  $a=1,.., \dbar$,  must satisfy a Clifford algebra,  and $M$ must anti-commute with all $\gamma_a$ in order
for the cross terms in $\CH^2$ to vanish: 
\beq
\label{CliffA}
\{ \gamma_a , \gamma_b \}  = 2 \delta_{ab}; ~~~~~\{ \gamma_a , M\}  = 0, ~~\forall a 
\eeq
Thus up to rescaling of $M$,  the set $\{\gamma_a , M \}$ form a Clifford algebra.   (The explicit form of 
$M$ will be given below,  where in general  it  will be an element of a  Clifford algebra times a constant   or 
tensored with an additional space.)

The conditions for $\bfP,  \bfT,  \bfC$ symmetry are the following $\forall a$:  
:
\barray
\label{Pcond}
&\bfP&:  ~~~~~\{  P, \gamma_a  \}  = 0 ,   ~~~~~~ \{ P, M \} = 0
\\
\label{Tcond}
&\bfT&:~~~~~T \gamma_a^T = - \gamma_a T,     ~~~~~T M^T = M T
\\
\label{Ccond}
&\bfC&:~~~~~C\gamma_a^T =  \gamma_a C ,       ~~~~~~~C M^T = -  M C
\earray
 The way these conditions are implemented is that one constructs $P, T, C$ satisfying the first condition in 
 each of the above cases,  which is the most stringent,   and then checks whether the second condition on $M$  is satisfied.

\def\timestimes{\otimes \cdots \otimes}

\section{Clifford algebra representation}  

In this section we describe an explicit representation of the Clifford algebra which we will utilize.  
A Clifford algebra is constructed from   $N$ basis elements $\Gamma_a$,  $a = 1,2, .... N$,    satisfying the relations:
\beq
\label{Cliffrel}
\{  \Gamma_a , \Gamma_b \}= 2 \delta_{ab}  
\eeq
We will refer to the  algebra generated by linear combinations of products of 
the $\Gamma_a$'s as the enveloping algebra of the Clifford algebra.  (In the mathematics literature, 
this enveloping algebra is simply referred to as the Clifford algebra.) 
 The degree of a monomial in the $\Gamma$'s is  the minimal number of factors subject to the relations
 (\ref{Cliffrel}).   
   Using the above relations of the basis elements,  the maximal degree of
an element of the enveloping algebra  is $N$  and the dimension of the enveloping  algebra,  i.e. the number of independent 
monomials,  is $2^N$.

Clifford algebras were classified abstractly  by Cartan.   They can all be realized as matrix algebras over the real or complex numbers,
or quaternions,  and possesses an 8-fold periodicity in the dimension of the basis.     
For our purposes,  it is more useful to work with an explicit representation.   
For $N=2n+1$ an  odd integer,  we can construct a representation of  the $N$  basis elements on a $2^n$ dimensional space 
in terms of an n-fold tensor product of Pauli matrices: 
\barray
\nonumber 
\Gamma_1 &=& \sigma_y \otimes  \sigma_z \timestimes \sigma_z  \\ 
\nonumber 
\Gamma_2 &=& \sigma_x \otimes  \sigma_z \timestimes \sigma_z  \\ 
\nonumber 
\Gamma_3 &=&  \one \otimes \sigma_y \otimes  \sigma_z \timestimes \sigma_z  \\ 
\nonumber 
\Gamma_4 &=&  \one \otimes \sigma_x \otimes  \sigma_z \timestimes \sigma_z  \\ 
&:&
\label{tensorsigma}
\\
\nonumber 
\Gamma_{2n-1} &=&  \one \timestimes \one \otimes \sigma_y
\\ 
\nonumber
\Gamma_{2n} &= &  \one \timestimes \one \otimes \sigma_x
\\
\nonumber
\Gamma_{2n+1}  & =&  \sigma_z \otimes \sigma_z \timestimes \sigma_z
\earray
where 
$\sigma_z =  \( \begin{smallmatrix} 1 & 0 \cr 0 & -1 \cr \end{smallmatrix} \)$, $\sigma_x =  \( \begin{smallmatrix} 0 & 1 \cr 1 & 0 \cr \end{smallmatrix} \)$, and $\sigma_y =  \( \begin{smallmatrix} 0 & -i \cr i  & 0  \cr \end{smallmatrix} \)$.
Note that all $\Gamma_a$ are  hermitian and  that $\Gamma_1  \Gamma_2 \cdots \Gamma_{2n+1} $ is
proportional to the identity.      The matrices $M_{ab} = \[ \Gamma_a , \Gamma_b \] /4i $ comprise  the  Lie algebra  for the irreducible 
spinor representation of SO($2n+1$).    The $a$ index of $\Gamma_a$ transforms as the vector representation of 
SO($2n+1$).   

Since the transpose is an anti-automorphism of the Clifford algebra,  and an involution,  
 then $A^T = \pm A$ for any monomial in the enveloping algebra.    We will need:
\barray
\nonumber 
\Gamma_a^T &=& - \Gamma_a ~~~{\rm if ~ a\neq 2n+1 ~ is ~ odd}   
\\
\label{transpose}
&=& ~\Gamma_a  ~~~~~{\rm if ~ a ~ is ~ even} 
\\
\nonumber
 \Gamma_{2n+1}^T &=& ~ \Gamma_{2n+1}
\earray
 
In the sequel,  the following elements   of the Clifford algebra  of degree $n$ and $n+1$   will play a central role in
constructing  the  $\bfT, \bfC$ symmetries: 
\beq
\label{Gs}
G = \Gamma_1 \Gamma_3 \Gamma_5 \cdots \Gamma_{2n-1} , ~~~~~   \Gtilde =  G \, \Gamma_{2n+1} 
\eeq
Using the transpose properties in eq. (\ref{transpose})  and the Clifford algebra relations,  one can show that they satisfy:
\beq
\label{Gtrans}
G^T =   (-1)^{n(n+1)/2}  G  , ~~~~~ \Gtilde^T = (-1)^{n(n-1)/2}  \Gtilde
\eeq 
We will also need:
\barray
\label{ssigns}
G \Gamma_{2n+1} &=&  (-1)^n \Gamma_{2n+1} G, ~~~
\Gtilde  \Gamma_{2n+1} = (-1)^n \Gamma_{2n+1} \Gtilde  
\\ \nonumber 
G \Gamma_{2n} &=& (-1)^n \Gamma_{2n} G, ~~~~~~
\Gtilde  \Gamma_{2n} = (-1)^{n +1} \Gamma_{2n} \Gtilde 
\earray
Finally,  note that $G \Gtilde \propto \Gamma_{2n+1}$,  which we will also need.  

\bigskip

\section{Generic Classification of Dirac fermions in any dimension}

It is useful to  first summarize the results of this section.  
Since the dimension of the above  enveloping algebra of the Clifford  algebra is $2^N = 2 \times 2^n \times 2^n$,  
any $2^n \times 2^n$ complex matrix can be expressed as an element of the enveloping  algebra.  
Thus,   the matrices $P, T, C$ can be expressed in terms of products of the $\Gamma_a$ matrices.  
We find that in a given dimension $d$,  there is a unique $T,C$ satisfying the stringent  first condition in  (\ref{Tcond}, \ref{Ccond}), 
which is either $G$ or $ \Gtilde$,  depending on the spatial dimension.      Which class this symmetry belongs to is determined by the transpose relations 
(\ref{Gtrans}).    The eight-fold periodicity arises from the even/odd properties of the powers in eq. (\ref{Gtrans}).
Namely,  $n(n-1)/2$ is even for $n=4m, 4m+1$ and odd for $n=4m+2, 4m+3$,  where $m$ is an integer.    On the other hand 
$n(n+1)/2$  is even for $n=4m, 4m+3$ and odd for $n=4m+1, 4m+2$.    Finally,  in order to obtain all $10$ classes, 
one needs to tensor in an additional space,  as will be explained.    We need to distinguish even and odd dimensions: 

\subsection{d odd}  

Let $d=2n+1$.    Without loss of generality one  can choose  $\gamma_a = \Gamma_a$ for $a=1, 2, ... 2n$, 
and $M = \Gamma_{2n+1}$,   since other choices are related by unitary SO(d) rotations.     
The $\bfP$ symmetry can be imposed with $P=\Gamma_{2n+1}$.

First consider $n$ odd.   Then the unique $T$ that satisfies the first condition in eq. (\ref{Tcond})  is 
$T=G$.    When $d=8m+3$, i.e. $n=4m+1$,  then $T^T = -T$.   In order to obtain $\bfT$ symmetry 
with the other sign in the  transpose,  one must tensor in an additional space.   Let $\vec{\tau}$  denote
another set of Pauli matrices.    Up to unitary transformations,  the additional factor in $T$ is either 
$1 $ or $i\tau_y$\cite{Dyson},  since they have opposite sign in the relation with their transpose. 
   Thus,  the other choice for $\bfT$ is $T' = i \tau_y \otimes G$,  
satisfying $T'^T = T'$.    On the other hand,  when $n=4m+3$, i.e. $d=8m+7$,  then  $T^T = T$ and $T'^T = -T'$.  
The $\bfC$ symmetry is similar.   The solution to the first eqn. in (\ref{Ccond}) is $C=\Gtilde$.   
If $d=8m+3$,  then $C^T = C$,  whereas if $d=8m+7$,  $C^T = -C$.   Again,  in order to obtain the other
sign in the transpose one needs to consider $C' = i\tau_y \otimes \Gtilde$.

Next consider $n$ even,  i.e. $n= 4m$ or $4m+2$,  corresponding to $d=8m+1, 8m+5$.   
 The symmetries are realized with  $T=\Gtilde$ and $C= G$.    For $d=8m+1$,     $T^T = -T$ and $C^T= C$,
 whereas for $d=8m+5$,  $T^T=T$ and $C^T = -C$.

These results can be summarized in the Table \ref{Table2}.   For a particular dimension $d$ modulo 8,  
the table indicates the ``primitive'' $T,C$,  and the specific sign in their transpose.    In each case,  in order
to obtain a representative with the opposite sign in the relation with their transpose,  one must use 
$T', C'$,  which henceforth will always denote the primitive $T,C$ tensored with $i\tau_y$.

\bigskip\bigskip

\subsection{d even}

Let $d=2n$.   It turns out one cannot construct a Clifford algebra on a space smaller than the 
$2^n$ dimensional space in eq. (\ref{tensorsigma}).   Thus  we take
$\gamma_a = \Gamma_a$ for $a = 1 ~ {\rm to }~ 2n-1$,  and $M=M^T= \Gamma_{2n}$.   
The extra matrix $\Gamma_{2n+1}$ commutes with all the SO(2n) generators,  thus
the $2^n$ dimensional space is irreducible,  and in fact the direct sum of the 2 spinor representations of
SO(d).     The  projectors onto these two representations are    $p_\pm = (1 \pm \Gamma_{2n+1})/2$, 
and we will refer to the projected representations as being of left or right handed chirality.   
Again $\bfP$ symmetry can be imposed with $P=\Gamma_{2n+1}$.

The construction of the $\bfT$, $\bfC$ symmetries is similar to the  odd d case.   For $d=8m$ and $8m+4$,
$T= \Gtilde$ and $C=G$,   whereas for $d=8m+2, 8m+6$,  they are reversed, i.e. $T=G, C=\Gtilde$.    
These results,  and the information on their transposes,  is also in  Table \ref{Table2}.   
Note that for the classes with both $\bfT, \bfC$ symmetry,  $P = T C^\dagger$ is proportional to 
$\Gamma_{2n+1}$,  consistent with our previous identification of $P$.
 
\def\p{$+1$}
\def\m{$-1$}

\begin{table}
\begin{center}
\begin{tabular}{|c|c|c|c|c|c|c|}
\hline\hline
$d$ mod 8  &  ~ ~~$T$~~ ~ &~~~ $T^T/T$~ ~~&~~~ $C$ ~ ~~   &   $C^T/C$ ~~~& $s_t$ ~~~& $s_c$~~~ \\
\hline\hline 
0     &     $\Gtilde$   &   $+1$    &     $G  $        &  $ +1$ &  \m  & \p  \\
1     &     $\Gtilde$   &   $+1$    &     $G  $          &  $ +1$ &\p  & \p   \\
2     &     $G         $   &   $-1$    &     $\Gtilde$   &  $ +1$ & \m &\p  \\
3     &     $G         $   &   $-1$    &     $\Gtilde$   &  $ +1$ &\m &\m   \\
4     &     $\Gtilde$   &   $-1$    &     $G  $   &  $ -1$  &\m &\p   \\
5     &     $\Gtilde$   &   $-1$    &     $G  $   &  $ -1$  &\p  &\p  \\
6     &     $G$            &   $+1$    &     $\Gtilde  $   &  $ -1$ &\m &\p   \\
7     &     $G$           &   $+1$    &     $\Gtilde  $   &  $ -1$ &  \m &\m  \\
\hline\hline 
\end{tabular}
\end{center}
\caption{\emph{  The implementation of $\bfT,  \bfC$ according to dimension.    The opposite sign of the relation
of $T,C$ to their transpose is realized with $T',C'$  (see text).  The signs $s_{t,c}$ are defined in section VII.   }}
\label{Table2}
\end{table}

\bigskip\bigskip

\section{Classification of Protected  Zero Modes and Topological Insulators}

In this section we classify gapless theories that are protected by the symmetries, i.e. the theories 
where the mass $M$ has a symmetry protected zero eigenvalue.     The existence of this zero mode
can arise in two ways:  either $M$ is forced to be zero,  or from the weaker condition $\det (M)=0$;   as explained below,  the first
way  corresponds to a $\Z$ or $2\Z$ topological insulator,  whereas the second is of type $\Z_2$.  

\subsection{AIII} 

The existence of TI's in class AIII in odd dimensions is easy to understand.   Recall that $\bfP$ symmetry is 
implemented with $\Gamma_{2n+1}$.    This leaves no $\Gamma$-matrix to associate with $M$ which 
necessarily  anti-commutes  with $P$.    Thus  $M$ is not allowed in odd dimensions in class AIII.

\subsection{Chiral classes in even dimensions}

Recall that in $d=2n$ even dimensions,  there are two additional $\Gamma$'s,  $\Gamma_{2n}$ and $\Gamma_{2n+1}$,   beyond the $d-1$ of them  associated with
the  $\gamma_a$'s.     This leads to the property of chirality.    In $d=2$ dimensions,  chirality corresponds to left or right movers
on the edge.   More generally we can define chiral states as follows.    Define projectors  $p_\pm = (1\pm g)/2$ onto 
states of ``left" verses ``right'' chirality,  where $g$ is either $\Gamma_{2n}$ or $\Gamma_{2n+1}$.   $M$ is then associated
with the other unused $\Gamma$,  e.g. if $g=\Gamma_{2n}$ then $M=\Gamma_{2n+1}$ and visa versa.  
A chiral theory is then defined as one  with a spectrum that consists of  only particles of right or left chirality.   
 
It is easy to see that $M$ necessarily couples both chiralities.    Using the fact that $M p_+ = p_- M$ and $p_+ p_- = 0$,   one 
has $\langle \psi |  M | \psi  \rangle =  \langle \psi | ( p_+ + p_-)   M  (p_+ + p_- )  | \psi \rangle  =  \langle \psi_L  |  M | \psi_R  \rangle 
+  \langle \psi_R  |  M | \psi_L  \rangle$.
     Thus a purely chiral theory has no possible mass term and is  a  candidate TI.

Chiral theories can have $\bfT$ or $\bfC$ symmetry,  but not both,  since $TC^\dagger$ is a $\bfP$ symmetry, 
and theories with $\bfP$ symmetry require both chiralities.   This is evident from the fact that $TC^\dagger
\propto  G \Gtilde  \propto  \Gamma_{2n+1}$,  which we have above associated with $P$.   
 Whether a chiral theory can have $\bfT$ or $\bfC$ symmetry depends on the dimension.   Let $S$ stand for either $T$ or $C$.
 The invariance of a chiral state under $S$ requires  $[S,p_\pm]=0$,   i.e. $[S, g] = 0$.   On the other hand,  if a mass is forbidden
 by the $S$ symmetry,   then this requires $\{ S,  M \} =0$.       One  then sees from eq. (\ref{ssigns}),  that $S$ then must be associated
 with $\Gtilde$.   According to Table \ref{Table2},    in dimensions $d=0,4$  this is a $\bfT$ symmetry,  whereas in $d=2,6$  it is a $\bfC$ symmetry.   
 This is consistent with the identifications made in \cite{EunAh} for $d=2$,  i.e. that left or right movers are invariant under $\bfC$ symmetry,
 whereas $\bfT$ symmetry exchanges them.     Since $M$ is forced to be zero,  these are TI of topological type $\Z$ (see the general discussion below).
 If the symmetry involves $T'$ or $C'$,  then the space is doubled,  and this should correspond to a TI  of type $2\Z$.  
 Thus,   for $d=0$,  there exist TI's in class  AII of type $\Z$ and in class AI of type $2\Z$.  
 whereas for $d=4$  chiral  TI's exist  in class  AI of type $\Z$ and in class AII of type $2\Z$.    It is a similar story  for TI's in classes
 C, D in $d=2,6$;   see Table \ref{PeriodicTable}.  
 
Finally  
one may consider a purely chiral theories with no $\bfT$ nor $\bfC$,  which are in class A.
For example,   for $d=2$,   chiral states do not preserve $\bfT$ since  $\bfT$ exchanges left and right movers.  
Thus any class that does not have $\bfT$ symmetry can be chiral,  namely C and D as described in the last paragraph,  but also A.     
Here a mass term is not allowed simply because the theory is chiral,   which should be distinguished from the above cases where  the mass term is 
also prohibited by $\bfT$ or $\bfC$ symmetry.    These are class A TI's of type $\Z$  in any even dimension.  
To summarize:

\bigskip
\n {\bf  $d=0$ mod 8.}    Chiral TI's of  type $\Z$ in classes A and AI,  and of type $2\Z$ in class AII.  

\bigskip
\n {\bf  $d=2$ mod 8.}    Chiral TI's of  type $\Z$ in classes A and D,  and of type $2\Z$ in class C.  

\bigskip
\n {\bf  $d=4$ mod 8.}    Chiral TI's of  type $\Z$ in classes A and AII  and of type $2\Z$ in class AI.  

\bigskip
\n {\bf  $d=6$ mod 8.}    Chiral TI's of  type $\Z$ in classes A and C,  and of type $2\Z$ in class D.  

\bigskip

\subsection{Non-chiral classes}

As described in section IV,  henceforth,   for odd dimensions we fix  $M=\Gamma_{2n+1}$  and for even dimensions $M=\Gamma_{2n}$.  
The latter is the natural choice since it is consistent with $P = T C^\dagger$ as explained above.  
The general form of $\bfT, \bfC$ are $T= \tau_t \otimes X_t$ and $C= \tau_c \otimes X_c$, 
where $X_{t,c}$ are either $G, \Gtilde$ according to Table \ref{Table2},  and $\tau_{t,c} = 1$ or $i\tau_y$.   
The ``mass'' can be generally expressed as $M= V \otimes \Gamma$,   where 
$\Gamma = \Gamma_{2n+1},  \Gamma_{2n}$ for $d=2n+1, 2n$ respectively.        We will consider only the minimal dimensions of
the space that $V$ lives in,  i.e. 1 or 2 dimensional.   
Let us define the signs $s_{t,c}$ as follows:    $X_{t,c} \Gamma = s_{t,c} \Gamma X_{t,c}$.     Then the constraints
on $V$ coming from $\bfT, \bfC$,  eq. (\ref{Tcond},\ref{Ccond}) are:
\beq
\label{Vcond}
\tau_t  V^T = s_t V \tau_t, ~~~~~~~~~\tau_c  V^T = -s_c  V \tau_c
\eeq
The signs $s_{t,c}$ follow from eq. (\ref{ssigns}).  
and are shown by dimension in Table \ref{Table2}. 

\def\det{{\rm det}}

\def\tauy{$ i \tau_y $}

\def\dim{{\rm dim}}

The symmetries constrain  $V$ according to eqs.  (\ref{Pcond},\ref{Tcond},\ref{Ccond}).    
A protected zero mode arises in one of two ways.   The symmetries can  force $V=0$,  which in lower
dimensions was associated with a $\Z$ topological invariant.   If the space is doubled, i.e. $\tau_{t,c} = i \tau_y$,   then this indicates 
the topological invariant is an even integer,  i.e. of type $2\Z$.    The other possibility is that the symmetries
lead to the condition $\det \, V= 0$ which implies $V$ has a zero eigenvalue.   As in $d=2,3$,  this condition
arises when a particular vector space has odd dimension,   and follows for example from 
$V^T = -V$,  which implies   $ \det V = - \det V = 0$;   this even/odd aspect  is associated with a 
$\Z_2$ topological insulator.   

Regardless of dimension, given the allowed $\tau_{t,c}$ and $s_{t,c}$,  one can identify 9 cases that have
a protected zero mode, and are  listed in Table \ref{Table3}.  
The two constraints one obtains besides $V=0$ are 
\barray
\label{detVodd}
V^T&=& -V ~~  \Longrightarrow  \det \, V = 0  ~~{\rm if}~  \dim (V)   {\rm ~  is~  odd}
\\
\nonumber
~ \\
\label{Vdiaga}
V &=&  \( \begin{matrix} a & 0 \cr 0 & -a \cr \end{matrix} \)  ~{\rm  with} ~ a^T = -a~  
\Longrightarrow \det V =  0 ~{\rm  if } ~ \dim(a)   ~ {\rm  is~} 1 
\earray

\begin{table}
\begin{center}
\begin{tabular}{|c||c|c|c|c|c|c|}
\hline\hline
case  &   ~~$\tau_t $~~  &~~ $\tau_c$  ~~&~~ $s_t$   ~~ & ~~ $s_c$ ~~  &  constraints on $V$  & type  \\
\hline\hline 
1      &   1    &     \none    &  \m    &   \none    &    eq. \ref{detVodd}  & $\Z_2$\\
\hline
2     &    \none    &     1    &  \none   &   \p    &      eq. \ref{detVodd}  & $\Z_2$  \\  \hline
3      &   1    &     1     &  \m    &   \p    &    eq. \ref{detVodd} & $\Z_2$  \\  \hline
4     &   1    &     1     &  \m    &   \m     &    $V=0$ & $\Z$  \\  \hline
5      &   1    &     1     &  \p    &   \p     &  $V=0$  & $\Z$  \\  \hline
6     &   \tauy    &     \tauy      &  \m    &   \m     &  $V=0$& $2\Z$  \\ \hline
7      &   \tauy     &     \tauy      &  \p    &   \p     & $V=0$  & $2\Z$   \\  \hline
8      &   \tauy     &     1      &  \p    &   \p    &  eq. \ref{Vdiaga}  & $\Z_2$  \\   \hline
9      &   1    &     \tauy      &  \m    &   \m     &  eq. \ref{Vdiaga} & $\Z_2$   \\
\hline\hline 
\end{tabular}
\end{center}
\caption{\emph{ The nine different ways a protected zero mode can arise, regardless of dimension. }}
\label{Table3}
\end{table}

TI's can now be classified by dimension as follows:   (i)  For a given dimension $d$,  identify $s_{t,c}$ from
Table \ref{Table2}.   (ii)  Identify which cases in Table \ref{Table3} apply for these values of $s_{t,c}$.   
(iii)   The transpose properties of $T, C$ can be inferred from Table \ref{Table2},  bearing in mind that
if $\bfT$ or $\bfC$ is $T'$ or $C'$,  then the sign is flipped.   (iv)  Identify the class using Table \ref{Table1}.
The results are the following:  

\vfill\eject

\n  {\bf  For d odd}

\n {\bf  $d=1$ mod 8.}     ~~ The cases from Table \ref{Table3} that apply are 2,5,7,8.   Examining their transpose properties in
Table \ref{Table2},
one sees that these correspond to TI's in classes  D,  BDI,  CII and DIII respectively.   

\bigskip

\n {\bf  $d=3$ mod 8.}    ~~ The cases from Table \ref{Table3} that apply are 1,4,6,9,  corresponding  to TI's in classes   AI,  DIII,  CI, CII   respectively.

\bigskip
\n { \bf   $d=5$ mod 8.}   ~~ The cases that apply are 2,5,7,8,   corresponding  to classes C, CII,  BDI and CI respectively.

\bigskip
\n {\bf  $d=7$ mod 8.}  ~~   The cases that  apply are 1,4,6,9,   corresponding  to TI's in classes  AI,  CI,  DIII and BDI  respectively.

\bigskip\bigskip 

\n  {\bf For  d even}  

In all even dimensions,  the cases from Table \ref{Table3} that apply are 1,2,3 and are thus all of type $\Z_2$.
    Which classes they belong to are again determined by the transpose properties
in Table \ref{Table2}  and comparing with Table \ref{Table1}:

\bigskip
\n { \bf   $d=0$ mod 8.}    ~~  Cases 1,2,3  give  TI's in classes  AI,  D and BDI  respectively, all of type $\Z_2$.   
The TI in class 
AI of type $\Z_2$ is new,  i.e. it was not in the original Periodic Table in\cite{Kitaev,Ryu}.

\bigskip
\n { \bf   $d=2$ mod 8.}    ~~  Cases 1,2,3  give  TI's in classes  AII,  D and DIII  respectively, all of type $\Z_2$.   
The TI in class 
D  of type $\Z_2$ is new.

\bigskip
\n { \bf   $d=4$ mod 8.}    ~~  Cases 1,2,3  give  TI's in classes  AII,  C and CII   respectively, all of type $\Z_2$.   
The TI in class 
AII   of type $\Z_2$ is new.

\bigskip
\n { \bf   $d=6$ mod 8.}    ~~  Cases 1,2,3  give  TI's in classes  AI,  C and CI  respectively, all of type $\Z_2$.   
The TI in class 
C   of type $\Z_2$ is new.

\bigskip

\section{The exceptionality of two dimensions}

For $d=2$ the above hamiltonian is   $\CH =  -i \sigma_y \d_x  +  V \sigma_x$,   where $M=V \sigma_x$.       Under a unitary transformation 
$\CH \to  U  \CH U^\dagger$,   with $U$ a rotation in the $\vec{\sigma}$ space  by $90^\circ$ about the $z$ axis  followed by a $90^\circ$ rotation about
the $x$ axis,   the hamiltonian is equivalent to 
\beq
\label{Ham2d}
\CH = -i \sigma_x  \d_x   +  V \sigma_z
\eeq
which is of the form studied in \cite{EunAh},  where in the latter $V=V_-$.    The reason more classes of TI's were found in \cite{EunAh},  namely 11, 
is the following.    In the generic classification of  the minimal Dirac hamiltonians in section V,  $T$ and $C$ were unique,  and 10 classes were obtained. 
In two dimensions,  the generic construction gives  $T= G = \sigma_y$ according to Table \ref{Table2}.      
However  since  $T$ only has to anti-commute with $\sigma_x$ for the hamiltonian (\ref{Ham2d}),   $\bfT$ can be realized also as $T= \sigma_z$. 
   These are gauge inequivalent since their transpose properties  are different.   
     A comprehensive classification of the most general  Dirac hamiltonians yields a richer,  more refined structure,  wherein 
some of the AZ classes have two inequivalent representatives\cite{EunAh}.    A total of 17 gauge-inequivalent classes were found, of which 11 
had protected zero modes and  conjectured to be topological insulators.

\section{Conclusions}

To summarize,   we have shown how the periodic table of topological insulators in all spatial dimensions can be
understood  in an alternative  manner,  namely by classifying symmetry protected zero modes of Dirac hamiltonians on the boundary.   
Our  holographic approach makes no use of topological invariants nor K-theory,  but is based only on generic properties of Clifford
algebras in any dimension.    Our analysis suggests an additional topological insulator of type $\Z_2$ in every even dimension.   
We also commented on why two dimensions has even more possible topological insulators\cite{EunAh}.

\section{Acknowledgments}
  AL would like to thank Csaba Csak and Eun-Ah Kim i for discussions. 
  This work is supported by the National Science Foundation under grant number  NSF-PHY-0757868 and by the ``Agence Nationale de la Recherche'' contract ANR-2010-BLANC-0414.

\def\mZ{$\Z$}
\def\Zt{$\Z_2$}
\def\ZZt{$\color{blue}{\Z}, \color{red}{\Z_2}$}
\def\s{}
\def\blu#1{\color{blue}{#1}}

\begin{table}
\begin{center}
\begin{tabular}{||c||cccccccc||}
\hline\hline
     ~      &        ~         &        ~      &        ~         &         $d ~ {\rm mod} ~ 8$         &        ~      &    ~    &       ~      &      ~            \\ 
\hline\hline 
 AZ class&        0       &        1          &        2            &         3                &        4            &    5             &         6          &      7           \\ %
 \hline\hline
 A      &        ~~\blu{\mZ}~~        &        ~~\none~~   &        ~~\blu{\mZ}~~        &        ~~ \none~~          &      ~~\blu{\mZ} ~~      &    ~~\none ~~   &      ~~\blu{ \mZ} ~~       &      ~~\none  ~~  \\ 
 AIII     &    \none    &           \mZ       &        \none         &        \mZ          &        \none       &    \mZ    &       \none       &      \mZ   \\ 
 AI     &        ~~~~ \ZZt ~~~~      &        \none       &        \none         &         \none          &       \blu{ 2\mZ  }    &    \none    &       \Zt      &      \Zt    \\ 
BDI     &        \Zt       &        \mZ       &        \none         &         \none          &        \none       &    2\mZ    &       \none       &      \Zt    \\ 
D      &        \Zt       &        \Zt       &       ~~~~\ZZt ~~~~   &         \none          &        \none       &    \none    &       \blu{2\mZ}        &      \none    \\ 
DIII       &      \none        &        \Zt      &        \Zt         &         \mZ           &        \none       &    \none    &       \none        &      2\mZ    \\ 
AII      &       \blu{2\mZ}         &        \none       &        \Zt        &         \Zt          &   ~~~~\ZZt~~~~        &    \none    &       \none       &      \none    \\ 
CII      &        \none        &        2\mZ       &        \none         &         \Zt         &        \Zt        &    \mZ      &       \none       &      \none    \\ 
C      &        \none        &        \none       &        \blu{2\mZ }        &         \none          &        \Zt        &    \Zt    &      ~~~~\ZZt~~~~   &      \none    \\ 
CI     &        \none        &        \none       &        \none         &         2\mZ           &        \none       &    \Zt     &       \Zt        &      \mZ     \\ 
\hline\hline 
\end{tabular}
\end{center}
\caption{\emph{ Periodic Table of topological insulators based on the classification of symmetry-protected zero modes.  
The chiral classes are all the ones of class A,  the first listed in cases with two entries, and those labeled $2\Z$ in even dimensions  (indicated in blue online).   The new candidate topological insulators 
are the second listed  in the  cases with two entries  (red online). }}
\label{PeriodicTable}
\end{table}

\end{document}